\documentclass[twocolumn,showpacs,preprintnumbers,amsmath,amssymb,prl]{revtex4}

\usepackage{graphicx}
\usepackage{dcolumn}
\usepackage{bm}
\begin{document}
\title{Hyperpolarizability effects in a Sr optical lattice clock}

\author{Anders Brusch}
\author{Rodolphe le Targat}
\author{Xavier Baillard}
\author{Mathilde Fouch\'e}
\author{Pierre Lemonde}

\email{pierre.lemonde@obspm.fr} \affiliation{LNE-SYRTE, Observatoire de Paris\\
61, Avenue de l'Observatoire, 75014, Paris, France}

\date{\today}

\begin{abstract}
We report the observation of the higher order frequency shift due
to the trapping field in a $^{87}$Sr optical lattice clock. We
show that at the magic wavelength of the lattice, where the first
order term cancels, the higher order shift will not constitute a
limitation to the fractional accuracy of the clock at a level of
$10^{-18}$. This result is achieved by operating the clock at very
high trapping intensity up to $400\,$kW/cm$^2$ and by a specific
study of the effect of the two two-photon transitions near the
magic wavelength.
\end{abstract}

\maketitle

The recent proposal and preliminary realizations of optical
lattice clocks open a promising route towards frequency standards
with a fractional accuracy better than
$10^{-17}$\,\cite{KatoPal03,Takamoto05,Ludlow05,Barber05}. A large
number of atoms are confined in micro-traps formed by the
interference pattern of laser beams which in principle allows both
the high signal to noise ratio of optical clocks with neutral
atoms\,\cite{Sterr04} and the cancellation of motional effects of
trapped ion devices\,\cite{Oskay05,Schneider05,Dube05,Margolis04}.
In contrast to the ion case an optical lattice clock requires high
trapping fields. The evaluation of their effects on the clock
transition is a major concern and is the subject of this letter.
For a Sr optical lattice clock, the typical requirement in terms
of trapping depth is about $10\,E_r$ with $E_r$ the recoil energy
associated to the absorption of a lattice
photon\,\cite{Lemonde05}. The corresponding frequency shift of
both clock states amounts to $36\,$kHz at 800\,nm, while a
relative accuracy goal of $10^{-18}$ implies a control of the
differential shift at the $0.5\,$mHz level, or $10^{-8}$ in
fractional units.

The frequency of the clock transition in a laser trap of depth
$U_0$ is shifted with respect to the unperturbed frequency $\nu_0$
according to
\begin{equation}
\nu =\nu_0 + \nu_1\frac{U_0}{E_r} + \nu_2\frac{U_0^2}{E_r^2} +
O\left(\frac{U_0^3}{E_r^3}\right), \label{eq:lightshift}
\end{equation}
with $\nu_1$ and $\nu_2$ proportional to the (dynamic)
polarizability and hyperpolarizability difference between both
states of the clock transition\,\cite{KatoPal03}. By principle of
the optical lattice clock, $\nu_1$ cancels when the laser which
forms the lattice is tuned to the "magic wavelength" $\lambda_m$.
Although this remains to be demonstrated experimentally, a control
of this first order term to better than $1\,$mHz seems
achievable\,\cite{KatoPal03}.

\begin{figure}
\begin{center}
\includegraphics[width=0.8\columnwidth]{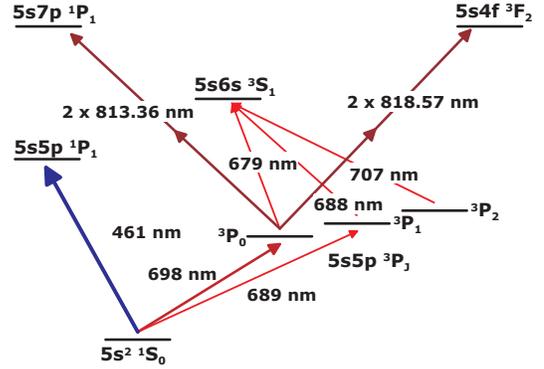}
\end{center}
\caption{Energy levels of Sr relevant to this paper. The clock
transition at 698\,nm couples the two lowest energy states of the
atom.} \label{fig:levels}
\end{figure}

The higher order term is a priori more problematic with no
expected cancellation. A theoretical calculation of the effect is
reported in Ref.\,\cite{KatoPal03} predicting a frequency shift of
$-2\,\mu$Hz$/E_r^2$ for a linear polarization of the lattice. The
calculation however was performed at the theoretical magic
wavelength of 800\,nm. The actual value\,\footnote{All wavelengths
throughout the paper are in vacuum.}, $\lambda_m=813.428(1)$\,nm
(see \cite{Takamoto05,Ludlow05} and below), lies near two
two-photon resonances which may considerably enhance the effect
and impede the realization of an accurate clock. The first one
couples $5s5p\, ^3P_0$ to $5s7p\, ^1P_1$ (Fig.\,\ref{fig:levels})
and is at a wavelength of $813.36$\,nm, or equivalently 30\,GHz
away from the magic wavelength. Although this $J=0\rightarrow J=1$
two-photon transition is forbidden to leading order for two
photons of identical frequencies\,\cite{Grynberg77}, it is so
close to the magic wavelength that it has to be a priori
considered. The second one resonantly couples $5s5p\, ^3P_0$ to
$5s4f\, ^3F_2$ at $818.57$\,nm and is fully allowed.

We report here an experimental study of higher order effects in a
$^{87}$Sr optical lattice clock operating at a very high trapping
depth up to $1400\,E_r$ and for a linear polarization of the
lattice. This depth is about a factor of ten higher than in the
other reported systems\,\cite{Takamoto05,Ludlow05} which enhances
the sensitivity to higher-order frequency shifts by two orders of
magnitude. At the maximum depth and for a detuning of the lattice
laser as small as 0.5\,GHz from the $^3P_0\rightarrow \, ^1P_1$
transition, higher order frequency shifts are below a few hertz.
On the other hand, when the lattice laser is tuned to within a few
hundred MHz of the $^3P_0\rightarrow \, ^3F_2$ transition,
non-linear frequency shifts of several kHz are observed. These
results, together with measurements at the magic wavelength, give
a stringent limit in optimal operating conditions. We show that
higher order frequency shifts will not limit the accuracy of a Sr
optical lattice clock down to the $10^{-18}$ level.

The high trapping depth is reached thanks to an enhancement
Fabry-P{\'e}rot cavity around the 1D vertical lattice. The circulating
power reaches 16\,W for a 650\,mW input at 813\,nm. The mode has a
waist radius of 90\,$\mu$m corresponding to a maximum
$U_0=1400\,E_r$ and an axial (resp. radial) oscillation frequency
of 260\,kHz (resp. 540\,Hz). An intracavity dichroic mirror
separates the lattice light from that which probes the clock
transition. It induces negligible intracavity loss at 813\,nm
while ensuring that less than $10^{-3}$ of the incident power at
698\,nm is reflected back to the atoms by the cavity mirror. The
lattice polarization is interferometrically filtered by an
intra-cavity quarter wave plate which lifts the degeneracy between
the linear polarization states parallel to the eigen axis of the
plate by half a free spectral range. The resulting polarization of
the trapping light is linear to better than $10^{-3}$. Its
wavelength $\lambda_L$ is controlled by means of a wave meter with
an accuracy of $10^{-3}$\,nm.

Atoms are loaded into the optical lattice from the magneto-optical
trap (MOT) described in\,\cite{Courtillot03OL}. Throughout the
loading cycle the lattice is overlapped with the MOT. Cold atoms
at the center of the trap are selectively optically pumped to the
metastable $^3P_{0,2}$ states by means of two "drain" lasers of
waist radius 50\,$\mu$m that are aligned to the lattice. They are
tuned to the $^1S_0 -\,^3P_1$ and $^3P_1-\,^3S_1$ transitions at
689 and 688 nm respectively (Fig.\,\ref{fig:levels}). Atoms in the
metastable states remain trapped provided their energy is smaller
than the $200\,\mu$K lattice depth. This leads to a continuous
loading of the lattice at a rate of about $10^5$ atoms/s. After
half a second of loading time the MOT and drain lasers are
switched off, and the atoms are repumped back to the ground state
using two lasers tuned to the $^3P_{0,2} -\,^3S_1$ transitions at
679 and 707\,nm. They are then cooled in the lattice to
$\sim\,10\,\mu$K in 50\,ms with the narrow $^1S_0-\,^3P_1$
intercombination line at 689\,nm\,\cite{Mukaiyama03}.

\begin{figure}
\begin{center}
\includegraphics[width=0.8\columnwidth]{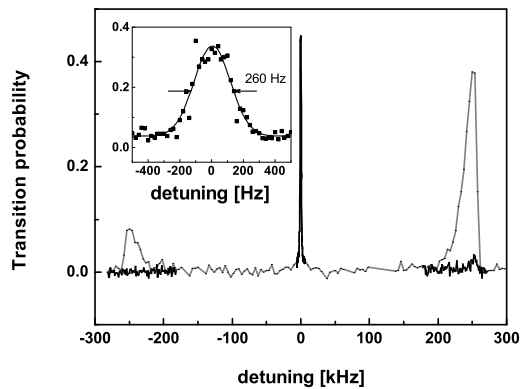}
\end{center}
\caption{Motional spectrum of the atoms in the optical lattice.
Each point corresponds to a single measurement. Black curve:
optimal operating conditions. The inset is a zoom on the carrier.
Grey curve: the motional sidebands are enhanced by applying long
probe pulses of 200\,ms at the maximum available power at
698\,nm.} \label{fig:sidebands}
\end{figure}

Following this preparation stage we probe the $^1S_0-\,^3P_0$
clock transition at 698\,nm. The frequency of the probe laser is
referenced to an ultra-stable cavity as described in
Ref.\,\cite{Quessada03}. The probe beam is aligned parallel to the
lattice and has a waist radius of 200\,$\mu$m. Its polarization is
linear and parallel to the lattice polarization. After the probe
pulse, the transition probability to $^3P_0$ is measured in a few
ms by laser induced fluorescence. A first detection pulse at
461\,nm gives the number of atoms remaining in the ground state
and ejects these atoms from the trap. Atoms in $^3P_0$ are then
repumped to the ground state and detected similarly. This method
allows a transition probability measurement which is insensitive
to the atom number fluctuations. The resonance is shown in
Fig.\,\ref{fig:sidebands}. The central feature (the carrier) is
zoomed in the inset of the figure. Its linewidth is 260\,Hz in
optimal operating conditions: $\lambda_L=\lambda_m$, probe pulse
of 10 ms duration and 2\,$\mu$W power. The resonance plotted here
was obtained at the maximum trapping depth and we observed no
clear dependence of its width and contrast with $U_0$.

The carrier is surrounded by motional sidebands shifted by the
oscillation frequency along the lattice direction. They are hardly
visible in optimal operation but can be enhanced by applying long
pulses of 200\,ms duration at the maximum available power
($2\,$mW) of the probe beam. The resulting spectrum is plotted in
grey in Fig.\,\ref{fig:sidebands}. The 1:5 ratio between the red
and blue sidebands show that 80\% of the atoms populate the
$|n_z=0\rangle$ motional state along the lattice axis,
corresponding to a temperature of $8\,\mu$K. The transverse
temperature is 10$\,\mu$K.

\begin{figure}
\begin{center}
\includegraphics[width=0.8\columnwidth]{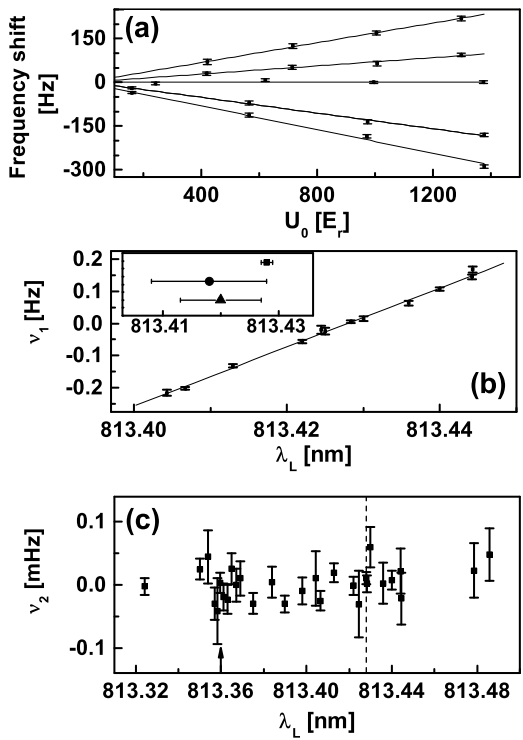}
\end{center}
\caption{(a): Frequency shift of the clock transition vs lattice
depth for 5 different lattice wavelength: 813.406 nm (four lowest
points in the graph), 813.413 nm, 813.428 nm, 813.436 nm and
813.444 nm (four highest points). Also shown is a linear fit of
each of these five sets of data. (b): First order frequency shift
vs $\lambda_L$ for $U_0=E_r$. Also plotted is a linear fit of
these data ($\chi^2=1.1$). The inset shows measurements of
$\lambda_m$. $\blacksquare$ : this work. {$\blacktriangle$} :
Ref.\,\cite{Takamoto05}. {\Large $\bullet$} :
Ref.\,\cite{Ludlow05}. (c): Higher order frequency shift vs
$\lambda_L$ for $U_0=E_r$. All these points are compatible with
zero. Their average is $-4(36)\times 10^{-7}\,$Hz ($\chi^2=1.02$).
The arrow on the $\lambda_L$ axis corresponds to the
$^3P_0\rightarrow\,^1P_1$ transition at 813.360\,nm. The vertical
dotted line is at the magic wavelength 813.428\,nm.}
\label{fig:higher_order_813}
\end{figure}

The effect of the trap on the clock transition is measured by
locking the frequency of the probe laser to the carrier for
various values of $U_0$ and $\lambda_L$. The trap depth $U_0$ is
adjusted to its desired value between $200\,E_r$ and $1400\,E_r$
by a linear ramp of 1\,ms duration between the cooling and probing
phases. This slightly decreases the transverse temperature while
the longitudinal motion is adiabatically cooled by following the
ramp. To lock the probe laser frequency we alternatively probe
both sides of the resonance and compute an error signal from the
difference between two successive measurements of the transition
probability. This also gives a measurement of the difference
between the atomic transition and the reference cavity frequency
which slowly fluctuates due to thermal effects. These fluctuations
behave essentially as a sine wave of peak to peak amplitude
300\,Hz and period 10 minutes. To derive the frequency shift due
to the lattice, we reject the cavity frequency fluctuations by a
factor 100 by a differential method\,\cite{Sortais00}. We
interleave measurements at 4 different lattice depths. We run the
clock for 19 cycles before $U_0$ is changed to the next of the
four interleaved values. The entire sequence is repeated typically
16 times. The cavity fluctuations are then modelled as a
polynomial which is determined by a least square fit of the
data\,\footnote{We use a tenth order polynomial fit of the data.
For any polynomial order between 5 and 14, the results are
unchanged to within a fraction of the error bars.}. The data are
corrected for the modelled cavity frequency fluctuations and
averaged. This yields 4 statistically independent
measurements\,\footnote{except for a common offset which doesn't
play any role in further data analysis.} of the clock transition
frequency versus $U_0$ with a standard deviation of about $5\,$Hz.
A typical set of such points is shown in Fig.
\ref{fig:higher_order_813}(a).

We perform a quadratic least square fit of each set of four points
which gives a measurement of the coefficients $\nu_1$ and $\nu_2$
of Eq.(\,\ref{eq:lightshift}). The coefficient $\nu_2$ is plotted
in Fig.\,\ref{fig:higher_order_813}(c) and found compatible with
zero for the whole range $[813.3\,$nm$,813.5\,$nm$]$ to within a
few tens of $\mu$Hz. A specially interesting wavelength region is
around $813.36$\,nm where the $^3P_0\rightarrow\,^1P_1$ two-photon
transition is expected. The contribution of a two-photon
transition to $\nu_2$ varies as $\Delta^{-1}$, with $\Delta$ the
detuning of the lattice with respect to the resonance. We magnify
this contribution by systematically spanning a frequency range of
$\pm 5\,$GHz with a 1\,GHz step around the expected value
(Fig.\,\ref{fig:higher_order_813}(c)). The null results of all
these measurements demonstrate that the higher order shift due to
the $^3P_0\rightarrow\,^1P_1$ transition is less than
$1\,\mu$Hz$/E_r^2$ for $\lambda_L=\lambda_m$. Despite its
proximity to the magic wavelength, this two-photon transition is
forbidden enough to not be a problem.

This set of experiments around $813.4\,$nm can also be used to
derive an accurate value of $\lambda_m$. Having shown that $\nu_2$
is negligible for $\lambda_L\sim\lambda_m$, better estimates of
$\nu_1$ are obtained with linear fits of each set of four points.
They are plotted in Fig.\,\ref{fig:higher_order_813}(b). We find
$\lambda_m=813.428(1)$\,nm in agreement with previously published
values\,\cite{Takamoto05,Ludlow05} as shown in the inset of
Fig.\,\ref{fig:higher_order_813}(b). The improvement by one order
of the accuracy of this measurement is a nice illustration of the
amplification of the effects of the lattice offered by a deep
trapping potential.

\begin{figure}
\begin{center}
\includegraphics[width=0.95\columnwidth]{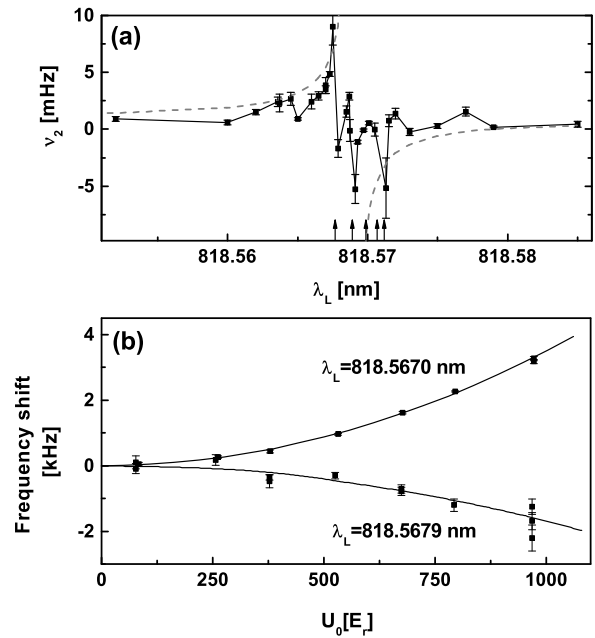}
\end{center}
\caption{(a): Higher order frequency shift around the
$^3P_0\rightarrow\,^3F_2$ transition at $818.570$\,nm for
$U_0=E_r$. The five vertical arrows on the wavelength axis
correspond to the hyperfine sub-states of $5s4f\, ^3F_2$ ($F=13/2$
to $F=5/2$ from left to right). (b): Atomic frequency shift vs
trapping depth for two lattice wavelengths on both sides of the
two-photon transition to sub-state $F=13/2$. The linear light
shift has been removed for clarity. The bold line is a fit of the
data with a parabola.} \label{fig:higher_order_2photon}
\end{figure}

We also studied the effect of the other two photon transition in
this wavelength region, the $^3P_0\rightarrow\, ^3F_2$ at
$818.57$\,nm. When tuned 5\,nm away from $\lambda_m$ we expect, in
addition to the effect of the two-photon coupling, a trivial
quadratic dependence of the atomic frequency vs $U_0$ due to the
imperfect cancellation of $\nu_1$ and to the inhomogeneity of the
laser intensity experienced by the atoms. We do observe a
substantial broadening and asymmetry of the resonance due to this
effect similar to what was reported in Ref.\,\cite{Takamoto03}.
The associated trivial quadratic frequency shift amounts to
$0.8\,$mHz$/E_r^2$ as measured several GHz away on both sides of
the two photon transition.

\begin{figure}
\begin{center}
\includegraphics[width=0.8\columnwidth]{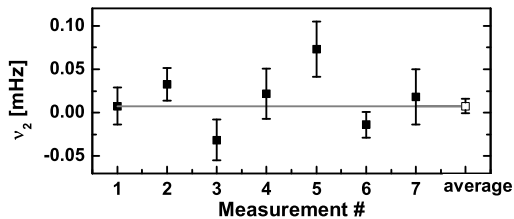}
\end{center}
\caption{Higher order frequency shift at the magic wavelength for
$U_0=E_r$. The value of the open square is the weighted average of
the seven other points ($\chi^2=1.8$).}
\label{fig:quadratic_magic}
\end{figure}

When tuned closer to the resonance, we clearly observe the
non-trivial quadratic frequency shift due to the two-photon
resonance itself. The effect is shown in
Fig.\,\ref{fig:higher_order_2photon}. Quadratic frequency shifts
of several kHz and changing sign depending on the side of the
transition being probed are observed as shown in
Fig.\,\ref{fig:higher_order_2photon}(b). This is a clear signature
of the higher order effects due to the particular transition under
investigation. When tuned exactly onto one of the five transitions
corresponding to the hyperfine structure of $^3F_2$, the lattice
laser induces severe loss (up to 90\%) of atoms when the 698\,nm
probe laser is tuned to resonance. This effect, which we attribute
to three-photon ionization from $^3P_0$, was used to determine the
position of the five hyperfine sub-states shown by arrows in
Fig.\,\ref{fig:higher_order_2photon}(a). The hyperfine structure
of $^3F_2$ leads to a complex dependence of the quadratic
frequency shift on the lattice wavelength around resonance. The
contribution of the five substates can interfere with each other,
which may be the cause of the oscillations of $\nu_2$ seen in
Fig.\,\ref{fig:higher_order_2photon}(a) on both sides of the
hyperfine manifold. We can deduce from our measurements a
conservative estimate of the contribution of the $^3P_0-\,^3F_2$
resonance to the higher order effects at $\lambda_L=\lambda_m$.
The grey dashed curved plotted in
Fig.\,\ref{fig:higher_order_2photon}(a) scales as the inverse of
the detuning of the lattice with respect to the center of gravity
of the hyperfine structure of $^3F_2$ and envelopes experimental
points. When extrapolated to the magic wavelength, it gives a
contribution to $\nu_2$ of $2\,\mu$Hz.

Finally we have performed an extensive series of measurements of
$\nu_2$ at $\lambda_L=\lambda_m$ resulting in the data plotted in
Fig.\,\ref{fig:quadratic_magic}. Their weighted average gives
$\nu_2(\lambda_m)=7(6)\,\mu$Hz. For a lattice depth of $10\,E_r$
the frequency shift is lower than 1\,mHz (one sigma) or $2\times
10^{-18}$ in fractional units. This demonstrates that the
frequency shift due to the atomic hyperpolarizability constitutes
no impediment to the accuracy of a Sr optical lattice clock down
to the $10^{-18}$ level. In addition, the effective laser
intensity seen by the atoms is certainly controllable at the
percent level\,\footnote{A control to within a few percents is
already achieved in our setup as evidenced by the nice alignement
of the measurements of $\nu_1$ shown in
Fig.\,\ref{fig:higher_order_813}(b).}. The performance of the
system would then be immune to higher order frequency shifts over
a broad lattice depth range, possibly up to $U_0=100\,E_r$. This
would provide a powerful lever for the experimental evaluation at
the $10^{-18}$ level of other effects associated for instance to
the dynamics of the atoms in the lattice or to cold collisions.
Collisions are expected negligible with polarized fermions, but
they have to be considered if one uses bosonic
isotopes\,\cite{Barber05}, such as $^{88}$Sr. By varying the
trapping depth, one can adjust the tunnelling rate and then
control the overlap of the wave functions of atoms confined in the
lattice, allowing the study of cold collisions in a new regime.

We thank A. Clairon and C. Salomon for useful comments and
discussions on the manuscript. SYRTE is Unit\'e Associ\'ee au CNRS
(UMR 8630) and a member of IFRAF (Institut Francilien de Recherche
sur les Atomes Froids). This work is supported by CNES.

\bibliographystyle{prsty}

\end{document}